\documentclass[superscriptaddress,prb,twocolumn,amsmath,amssymb]{revtex4} %twocolumn,
\usepackage{graphicx,natbib}

\newcommand{\RM }[1]{\mathrm{#1}}

\def\kB{ k_{\RM{B}} }

\def\DRosen{{D\rho^{1/3}(\kB T/m_i)^{-1/2}}}
\newcommand{\DRoseni}[1][i]{{D_{#1}\rho^{1/3}(\kB T/m_i)^{-1/2}}}

\def\stwo{{s^{(2)}}}

\newcommand{\stwoi}[1][i]{{s^{(2)}_{#1}}}

\def\Dr{{D^R}}
\newcommand{\Dri}[1][i]{{D_{#1}^R}}

\def\naive{{na\"{i}ve }}

\begin{document}

\title{Composition and concentration anomalies for structure and
  dynamics of Gaussian-core mixtures}

\author{Mark J. Pond} \email{mjp736@che.utexas.edu} \author{William P.
  Krekelberg} \affiliation{Department of Chemical Engineering,
  University of Texas at Austin, Austin, TX 78712.}

\author{Vincent K. Shen} \email{vincent.shen@nist.gov}
\affiliation{Physical and Chemical Properties Division, National
  Institute of Standards and Technology, Gaithersburg, Maryland
  02899-8380, USA}

\author{Jeffrey R. Errington} \email{jerring@buffalo.edu}
\affiliation{Department of Chemical and Biological Engineering,
  University at Buffalo, The State University of New York, Buffalo,
  New York 14260-4200, USA}

\author{Thomas M. Truskett} \email{truskett@che.utexas.edu}
\thanks{Corresponding Author} \affiliation{Department of Chemical
  Engineering, University of Texas at Austin, Austin, TX 78712.}
\affiliation{Institute for Theoretical Chemistry, University of Texas
  at Austin, Austin, TX 78712.}

\begin{abstract}
  We report molecular dynamics simulation results for two-component
  fluid mixtures of Gaussian-core particles, focusing on how tracer
  diffusivities and static pair correlations depend on temperature,
  particle concentration, and composition.  At low particle
  concentrations, these systems behave like simple atomic mixtures.
  However, for intermediate concentrations, the single-particle
  dynamics of the two species largely decouple, giving rise to the
  following anomalous trends.  Increasing either the concentration of
  the fluid (at fixed composition) or the mole fraction of the larger
  particles (at fixed particle concentration) enhances the tracer
  diffusivity of the larger particles, but decreases that of the
  smaller particles.  In fact, at sufficiently high particle
  concentrations, the larger particles exhibit higher mobility than
  the smaller particles.  Each of these dynamic behaviors is
  accompanied by a corresponding structural trend that characterizes
  how either concentration or composition affects the strength of the
  static pair correlations.  Specifically, the dynamic trends observed
  here are consistent with a single empirical scaling law that relates
  an appropriately normalized tracer diffusivity to its
  pair-correlation contribution to the excess entropy.
\end{abstract}
\maketitle

Fluids of identical particles interacting via the Gaussian-core (GC)
pair potential have been the subject of many recent
investigations.\cite{Stillinger1997Negative-therma,
  Lang2000Fluid-and-solid, Louis2000Mean-field-flui,
  Prestipino2005Phase-diagram-o-JCP, Giaquinta2005Re-entrant-Melt,
  Mausbach2006Static-and-dyna, Zachary2008Gaussian-core-m,
  Wensink2008Long-time-self-, Krekelberg2009Gaussian-dynamics}
Continued interest in this model system, introduced by Stillinger in
1976,\cite{Stillinger1976Phase-transitio} can be attributed in part to
the fact that the GC potential is a simple and computationally
tractable idealization of the soft, effective interparticle repulsions
that can exist between large molecular species (e.g., star polymers)
or self-assembled structures (e.g., micelles) in
solution.\cite{Likos2001effective-interactions} The GC fluid is also a
compelling model to study because it exhibits several unusual physical
properties that are typically associated with molecular or
complex-fluid systems with more complicated interactions.  For
example, at low temperature, the GC fluid displays a re-entrant
freezing transition \cite{Prestipino2005Phase-diagram-o-JCP,
  Giaquinta2005Re-entrant-Melt, Zachary2008Gaussian-core-m,
  Stillinger1976Phase-transitio} negative thermal
expansivity,\cite{Stillinger1997Negative-therma,
  Stillinger1978Study-of-meltin} and its isothermal compressibility
increases upon isobaric cooling.\cite{Mausbach2006Static-and-dyna}
Although the structural and dynamic properties of the GC fluid are
qualitatively similar to those of simpler fluids at low particle
concentrations, they become anomalous at sufficiently high particle
concentrations.  For example, the single-particle dynamics, quantified
by, e.g., the self diffusivity, become progressively faster upon
increasing particle concentration (diffusivity anomaly).
\cite{Mausbach2006Static-and-dyna, Wensink2008Long-time-self-,
  Krekelberg2009Gaussian-dynamics, Stillinger1978Study-of-meltin} The
static pair correlations,\cite{Lang2000Fluid-and-solid,
  Louis2000Mean-field-flui, Giaquinta2005Re-entrant-Melt,
  Mausbach2006Static-and-dyna, Zachary2008Gaussian-core-m,
  Wensink2008Long-time-self-, Krekelberg2009Gaussian-dynamics,
  Stillinger1976Phase-transitio} quantified by, e.g., the two-body
excess entropy~$\stwo$, also weaken upon increasing particle
concentration (structural
anomaly).\cite{Krekelberg2009Gaussian-dynamics}

The differences between the structural behavior of the GC fluid at low
versus high particle concentration can be qualitatively understood by
considering the Gaussian form of the repulsion.  At low concentration
and low temperature, the average interparticle separation is larger
than the range of the interaction.  Thus, the part of the GC potential
that the particles typically sample when they ``collide'' is steeply
repulsive.  Under these conditions, small increases in concentration
lead to the build up of short-range static correlations (i.e.,
coordination shell structure), similar to what occurs in the
hard-sphere fluid.\cite{Lang2000Fluid-and-solid,
  Louis2000Mean-field-flui, Giaquinta2005Re-entrant-Melt,
  Mausbach2006Static-and-dyna, Zachary2008Gaussian-core-m,
  Wensink2008Long-time-self-, Krekelberg2009Gaussian-dynamics,
  Stillinger1976Phase-transitio} However, at sufficiently high
concentration, the bounded form of the GC potential allows the average
interparticle separation to become much smaller than the range of the
interaction.  As a result, particles are effectively penetrable and
constantly experience soft repulsive forces from many neighbors.
These forces largely cancel one another, creating a ``mean field''.
Further increasing the concentration only makes this effect more
pronounced, paradoxically driving the high-density system toward an
ideal-gas-like structure.  \cite{Lang2000Fluid-and-solid,
  Louis2000Mean-field-flui, Giaquinta2005Re-entrant-Melt,
  Mausbach2006Static-and-dyna, Zachary2008Gaussian-core-m,
  Wensink2008Long-time-self-, Krekelberg2009Gaussian-dynamics,
  Stillinger1976Phase-transitio}

Less is understood about the microscopic origins of the anomalous
relationship between diffusivity and particle concentration, although
the results of recent investigations indicate that the unusual
dynamical trends are closely linked to the aforementioned structural
anomalies.\cite{Wensink2008Long-time-self-,
  Krekelberg2009Gaussian-dynamics} In particular, the equilibrium GC
fluid exhibits a semi-quantitative scaling
relation\cite{Krekelberg2009Gaussian-dynamics} between self
diffusivity~$D$ and the two-body excess entropy~$\stwo$.
Interestingly, this relationship is ``normal'' in the sense that it is
similar to that observed for a wide variety of simpler fluids that do
not exhibit either structural or dynamic anomalies.
\cite{Krekelberg2009Gaussian-dynamics, Rosenfeld1977Relation-betwee,
  Rosenfeld1999A-quasi-univers,
  Dzugutov1996A-univeral-scal,Mittal2007Relationships-b} Stated
differently, the diffusivity anomaly of the equilibrium GC fluid
disappears when one plots $D$ versus $\stwo$ instead of particle
concentration.\cite{Krekelberg2009Gaussian-dynamics} Similar trends
have also recently been reported for other equilibrium fluids with
dynamic and structural anomalies, e.g., models with water-like
interactions
\cite{Errington2001Relationship-be,Mittal2006Relationship-be,
  Errington2006Excess-entropy-} or colloid-like, short-range
attractions.  \cite{Mittal2006Quantitative-Li,
  Krekelberg2007How-short-range}

In this paper, we further explore the relationship between structure
and dynamics in simple models for complex fluids by studying, via
molecular simulation, binary mixtures of GC particles.  The fluid
phase behavior of these systems has already been studied extensively.
\cite{Archer2001Binary-Gaussian} However, here we present, to our
knowledge, the first investigation of the relationships between the
static pair correlations of the fluid and the tracer diffusivities of
the two components.  

Specifically, we study the following questions
about how these quantities depend on particle concentration and
mixture composition.  
Are the trends in the tracer diffusivities of the two components of
the GC mixture closely coupled?  Do they scale in a simple way with a
single measure of the overall strength of the pair correlations (e.g.,
$\stwo$)?  Or, alternatively, is there a significant decoupling of the
single-particle dynamics of the two species?  If this latter scenario
holds, do the resulting trends in tracer diffusivities track
decoupled, species-dependent measures of static structure? Finally,
what are the implications of the answers to the above for the
compositional dependencies of structural order and tracer diffusivity
at low versus high particle concentration?

To address these questions, we use molecular dynamics simulations to
investigate equilibrium two-component fluid mixtures of particles that
interact via pair potentials of the GC form,
$\phi_{ij}(r)=\epsilon_{ij} \exp[-(r/\sigma_{ij})^2]$.  Here, $r$ is
the interparticle separation, and the parameters $\epsilon_{ij}$ and
$\sigma_{ij}$ characterize the energy scale and range of the
interactions, respectively, between particles of type $i$ and $j$ with
$i,j \in \{A,B\}$.  Since we want to understand the behavior of
uniform binary fluids, we assign numerical values to the parameters
that favor mixing.  Specifically, we adopt a
set\cite{Archer2001Binary-Gaussian} introduced earlier
($\sigma_{BB}=0.665 \sigma_{AA}$;
$\sigma_{AB}=(0.5[\sigma_{AA}^2+\sigma_{BB}^2])^{0.5}$;
$\epsilon_{AA}=\epsilon_{BB}$; $\epsilon_{AB}=0.944 \epsilon_{AA}$),
in which the $\sigma_{ij}$ were chosen to mimic binary mixtures of
self-avoiding polymers in solution.  \cite{Louis2000Mean-field-flui}
We truncate all pair potentials at an interparticle separation of
3.2$\sigma_{AA}$, and treat the particles of the two species as having
equal masses ($m_A=m_B$).

We carry out the simulations in the microcanonical ensemble,
numerically integrating Newton's equations of motion with the
velocity-Verlet scheme~\cite{Allen1987Computer-Simula} using a time
step of 0.05$\sqrt{m_A \sigma_{AA}^2 / \epsilon_{AA}}$.  We use
$N=3000$ GC particles and a periodically replicated simulation cell,
the volume $V$ of which is chosen to realize specific values of
reduced total concentrations (i.e., particle densities) in the range
$0.05 \le \rho \sigma_{AA}^3 \le 1$, where $\rho=N/V$.  We investigate
mixtures over a wide range of composition ($0.1 \le x_A \le 0.9$,
where $x_A$ is the mole fraction of species A) and reduced temperature
($0.05 \le k_BT/\epsilon_{AA} \le 0.40$).  To characterize the
single-particle dynamics of species $i$, we compute its tracer
diffusivity $D_i$ by fitting the long-time ($t \rightarrow \infty$)
behavior of its average mean-squared displacement $\langle {\Delta
  r_i}^2 \rangle$ to the Einstein formula, $D_i = \langle {\Delta
  r_i}^2 \rangle / 6t$.\cite{note-error-est}  % \footnote{We use Student's t distribution
%   together with the tracer diffusivities from five independent runs to
%   estimate 95\% confidence intervals for the $D_i$.}

We compute the two-body excess entropy~$\stwo$ directly from the
partial radial distribution functions~$g_{ij}(r)$ of the fluid using
the
expression,\cite{Hernando1990Thermodynamic-pot,Samanta2001Universal-Scali}
\begin{equation}
  \label{eq:s2}
  \stwo=\sum_i x_i \stwoi
\end{equation}
where $\stwoi[i]$ is given by
\begin{equation}
  \label{s2ij}
  \frac{\stwoi[i]}{\kB} = - \sum_j  \frac{x_j \rho}{2} \int
  [g_{ij}({\bf r}) \ln   g_{ij}({\bf{r}})-g_{ij}({\bf{r}})+1]
  d{\bf{r}}
\end{equation}
Both $-\stwo$ and $-\stwoi[i]$ are non-negative and can be viewed as
translational structural order
metrics.\cite{Truskett2000Towards-a-quant} The former characterizes
the overall strength of the pair correlations in the
fluid,\cite{Truskett2000Towards-a-quant} while the latter quantifies
the amount of pair structuring surrounding particles of type $i$.

\begin{figure}[t]
  \includegraphics[clip]{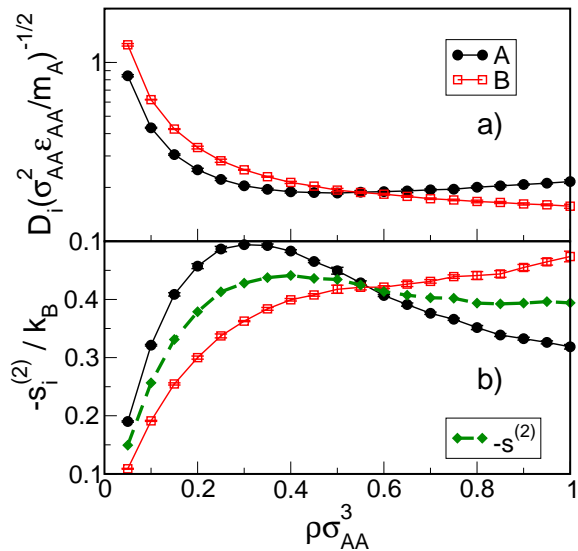}
  \caption{(a)Tracer diffusivity $D_i$ and (b) structural order metric
    -$\stwoi[i]$, with $i \in \{A,B\}$, versus concentration $\rho
    \sigma_{AA}^3$ for the binary Gaussian-core fluid mixture
    discussed in the text.  The collective structural order metric
    -$\stwo$ is also included in (b).  The temperature is $\kB T/
    \epsilon_{AA}=0.1$ and the mole fraction is $x_A=0.5$.}
  \label{fig:denscomps}
\end{figure}

The first issue that we investigate using our simulation data is how
closely the single-particle dynamics of the two species are coupled.
Figure~\ref{fig:denscomps}(a) shows how the computed tracer
diffusivities, $D_A$ and $D_B$, depend on density $\rho \sigma_{AA}^3$
for an equimolar ($x_A=0.5$) mixture at a temperature of $\kB
T/\epsilon_{AA}=0.1$.  As can be seen, $D_A$ follows the same type of
non-monotonic trend observed for the self diffusivity of the
single-component GC fluid,\cite{Krekelberg2009Gaussian-dynamics}
displaying an anomalous dependency on particle concentration
[$(\partial D_A / \partial \rho)_{T,x_A}>0$] for densities greater
than $\rho \sigma_{AA}^3 \approx 0.4$. On the other hand, $D_B$ shows
behavior consistent with that of simple fluids, monotonically
decreasing with $\rho \sigma_{AA}^3$ over the density range examined
here.  The fact that $D_A$ and $D_B$ decouple in this way gives rise
to a dynamic crossover density, above which the larger A particles
exhibit higher mobility than the smaller B particles.  It also
suggests that one cannot trivially correlate both the $D_A$ and $D_B$
trends with a single, collective measure of structural order for the
fluid, such as $\stwo$ [see, e.g., Fig.~\ref{fig:denscomps}(b)], a
point we examine further below.

Does increasing particle concentration result in a corresponding
decoupling of species-specific structural metrics that, in turn,
correlate with the dynamical trends of A and B particles?  To examine
this possibility, we first plot in Fig.~\ref{fig:denscomps}(b) the
density dependencies of $-\stwoi[A]$ and $-\stwoi[B]$.  As can be
seen, there is indeed a structural decoupling.  While both
$-\stwoi[A]$ and $-\stwoi[B]$ increase with density at low values of
$\rho \sigma_{AA}^3$, the increase in ordering is initially more
pronounced for the structure surrounding the A particles.  This is to
be expected because A particles have larger effective contact
diameters with their neighbors, and thus respond by building up
stronger static pair correlations at low density.  However, the large
size of the A particles, coupled with the bounded nature of the GC
interaction, means that A particles are also forced into more
interparticle overlaps than the B particles as the density is
increased.  This ultimately leads to a weakening of the static
correlations (i.e., coordination shell structure) of the A particles,
and hence a maximum in $-\stwoi[A]$ at an intermediate value of $\rho
\sigma_{AA}^3$.  As should be expected based on the smaller size of
the B particles, a maximum in $-\stwoi[B]$ (and a corresponding
minimum in $D_B$) can also occur at significantly higher densities, if
phase separation of the mixture does not occur first.\cite{note-anomalies}% \footnote{In the
%   limit of pure B ($x_A=0$), one
%   observes\cite{Krekelberg2009Gaussian-dynamics} both $(\partial D_B /
%   \partial \rho)_{T,x_A}>0$ and $(\partial \stwoi[B] / \partial
%   \rho)_{T,x_A}>0$ for reduced densities greater than $\rho
%   \sigma_{AA}^3 \approx 1.4$.  However, we have found in our studies
%   of this system that, for non-zero values of $x_A$, the binary GC
%   fluid will often phase separate at densities below where the onset
%   of anomalous dynamic and structural behavior occurs.}
Interestingly, similar to what is observed for the tracer
diffusivities, one of the consequences of the $\stwoi$ decoupling
described above is the presence of a structural crossover density
above which the smaller B particles exhibit stronger pair
correlations than the larger A particles.

The data in Fig.~\ref{fig:denscomps}(a) and (b) suggests a negative
correlation between $D_A$ and $-\stwoi[A]$ (and also between $D_B$ and
$-\stwoi[B]$), in which the structural crossover
($\stwoi[A]=\stwoi[B]$) occurs at approximately the same density as
the dynamic crossover ($D_A=D_B$).  In fact, although we focus on one
particular pairing of composition and temperature in
Fig.~\ref{fig:denscomps}, these trends are exhibited by this system
for a wide range of compositions and temperatures
% (see Fig. 1 in the
% EPAPS supplementary material at [URL will be inserted by
% AIP]).
(see the supplementary Fig. \ref{fig:denscompsup})

\begin{figure}
  \includegraphics[clip]{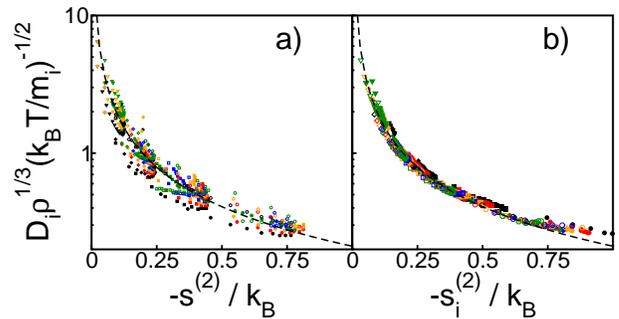}
  \caption{Rosenfeld scaled tracer diffusivity $\Dri = \DRoseni$
    versus (a) two body excess entropy -$\stwo$ and (b) its contribution
    from structuring around type $i$ particles -$\stwoi$, with $i \in
    \{A,B\}$, for the binary Gaussian-core fluid mixture discussed in
    the text. Shown are mole fractions $x_A= 0.1$ (black), $0.3$
    (red), $0.5$ (orange), $0.7$ (blue), and $0.9$ (green) and
    temperatures $\kB T/\epsilon_{AA}= 0.05$ ($\circ$), $0.1$
    ($\square$), $0.2$ ($\diamond$) and $0.4$ ($\triangledown$).
    Filled and open shapes represent A and B particles, respectively.
    The dashed line indicates a least-square fit to a power law
    relationship for the single-component GC
    fluid\cite{Krekelberg2009Gaussian-dynamics}, $\Dr = 0.208
    \stwo^{-0.972}$.}
  \label{fig:drivss2}
\end{figure}

\begin{figure}
  \includegraphics[clip]{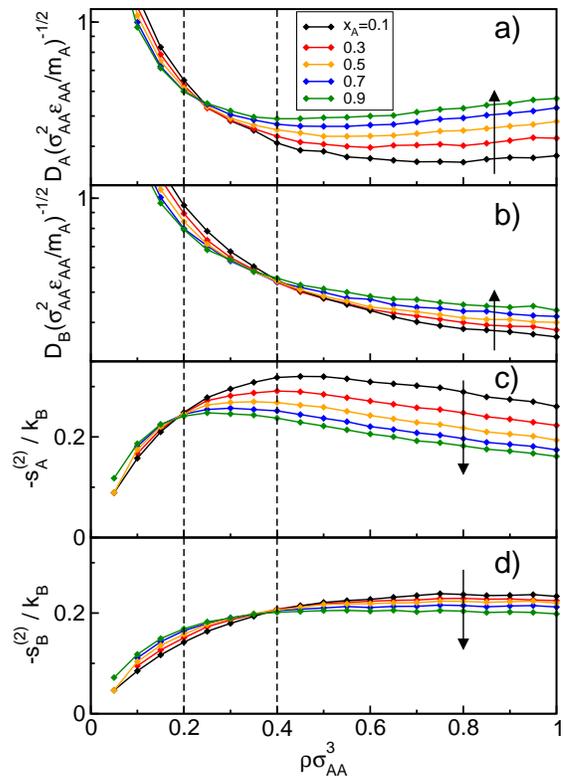}
  \caption{Tracer diffusivity $D_i$ for (a) A particles and (b) B
    particles and structural order metric -$\stwoi[i]$ for (c) A
    particles and (d) B particles for the binary Gaussian-core mixture
    discussed in the text.  All are plotted versus density $\rho
    \sigma_{AA}^3$. Data shown is for temperature $\kB
    T/\epsilon_{AA}=0.2$ and mole fractions $x_A=0.1$, $0.3$, $0.5$,
    $0.7$ and $0.9$. Arrows indicate increasing $x_A$. The dashed
    lines separate the approximate low, intermediate and high density
    ranges described in the text.}
  \label{fig:fullT02}
\end{figure}

To quantitatively examine the correlation between single-particle
dynamics and structure, we studied possible generalizations of a
scaling relationship that describes the behavior of the
single-component GC fluid.  In the case of the single-component fluid,
the so-called Rosenfeld scaled\cite{Rosenfeld1999A-quasi-univers} self
diffusivity $\Dr = \DRosen$ is approximately a single-valued function
of $-\stwo$ across a wide range of temperature and
density.\cite{Krekelberg2009Gaussian-dynamics} Figure
\ref{fig:drivss2}(a), however, shows that a \naive extension of this
result for mixtures, $\Dri = \DRoseni$ versus $-\stwo$, does not
adequately collapse the data for either of the two species.  This
should not be particularly surprising, given the dynamical and
structural decouplings shown in Fig.~\ref{fig:denscomps}(a) and (b).

On the other hand, Figure \ref{fig:drivss2}(b) examines a
species-specific extension of the single-component scaling law, $\Dri$
versus $-\stwoi$.  Interestingly, not only does this generalization
collapse the temperature, density, and compositional dependencies of
tracer diffusivity for each individual particle type, but the
behaviors of the two species are, to a good approximation, accounted
for by the mathematical form of the scaling law for the
single-component GC fluid.

We now explore what this correlation implies about 
how mixture composition affects the tracer diffusivities [Figure
\ref{fig:fullT02}(a),(b)] and species-specific pair-correlation
contributions to excess entropy [Figure \ref{fig:fullT02}(c),(d)].  
As should be expected, at low values of
density ($\rho \sigma_{AA}^3 < 0.2$), the response of the system to
changes in composition is normal, i.e.,
qualitatively similar to that of simple atomic or hard-sphere-like
mixtures where interparticle ``collisions'' dominate.  
Under these conditions, increasing the mole fraction of
the larger A particles effectively increases the ``packing
fraction'' of the fluid, which in turn decreases the mobility and
increases the local structural order surrounding both types of 
particles.   

At high values of density, on the other hand, the fact that the
bounded GC potential allows for significant interparticle overlaps
changes the physics.  
Since increasing the mole fraction of the larger particles (at constant
density) here increases the number of overlaps and nudges the system
toward the mean-field fluid, one expects anomalous behavior in
dynamics and structure, i.e. a
corresponding increase in $D_i$ and decrease in $-\stwoi[i]$ for both 
species.  
Indeed, Figure \ref{fig:fullT02} shows that
these anomalous compositional trends for dynamics and structure 
do occur for $\rho \sigma_{AA}^3 > 0.4$.

In closing, we note a final manifestation 
of the decoupled behavior for the two species of this GC mixture.  
Specifically,
the densities at which the compositional trends for structure and
dynamics transition from normal to anomalous are significantly different
for the two species, with the larger species logically becoming
anomalous at a lower overall density.  The implication is that there
is a fairly wide range of intermediate fluid densities (approximately $0.2 < \rho \sigma_{AA}^3 < 0.4)$
for which the structure and dynamics of the A particles behave
anomalously, while those of the B particles
behave normally, with respect to changes in composition.

Two authors (T.M.T and J.R.E) acknowledge financial support of the
National Science Foundation (CTS-0448721 and CTS-028772,
respectively). One author T.M.T. also acknowledges support of the
Welch Foundation (F-1696) and the David and Lucile Packard Foundation.
M.J.P. acknowledges the support of the Thrust 2000 - Harry P.
Whitworth Endowed Graduate Fellowship in Engineering. The Texas
Advanced Computing Center (TACC) provided computational resources for
this study.

\begin{figure*}
  \includegraphics[clip]{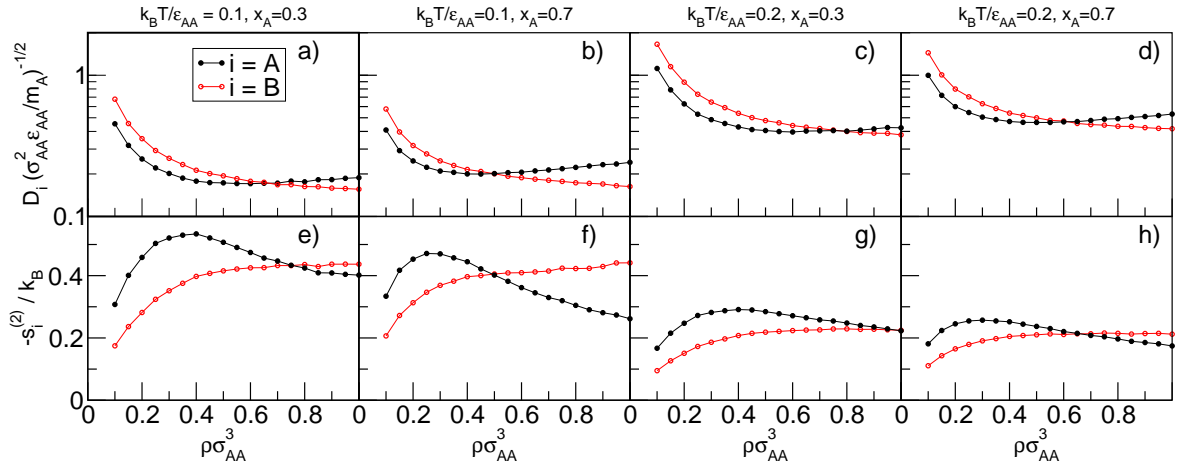}
  \caption{(a-d)Tracer diffusivity $D_i$ and (e-h) structural order
    metric -$\stwoi[i]$, with $i \in \{A,B\}$, versus
    concentration $\rho \sigma_{AA}^3$ for the binary Gaussian-core
    fluid mixture discussed in the text. The columns are systems at
    different temperatures and mole fractions, which are listed
    above the respective column.}
  \label{fig:denscompsup}
\end{figure*}

%\bibliography{mjp_firstbib,notes.bib}

\begin{thebibliography}{28}
\expandafter\ifx\csname natexlab\endcsname\relax\def\natexlab#1{#1}\fi
\expandafter\ifx\csname bibnamefont\endcsname\relax
  \def\bibnamefont#1{#1}\fi
\expandafter\ifx\csname bibfnamefont\endcsname\relax
  \def\bibfnamefont#1{#1}\fi
\expandafter\ifx\csname citenamefont\endcsname\relax
  \def\citenamefont#1{#1}\fi
\expandafter\ifx\csname url\endcsname\relax
  \def\url#1{\texttt{#1}}\fi
\expandafter\ifx\csname urlprefix\endcsname\relax\def\urlprefix{URL }\fi
\providecommand{\bibinfo}[2]{#2}
\providecommand{\eprint}[2][]{\url{#2}}

\bibitem[{\citenamefont{Stillinger and
  Stillinger}(1997)}]{Stillinger1997Negative-therma}
\bibinfo{author}{\bibfnamefont{F.~H.} \bibnamefont{Stillinger}}
  \bibnamefont{and} \bibinfo{author}{\bibfnamefont{D.~K.}
  \bibnamefont{Stillinger}}, \bibinfo{journal}{Physica A}
  \textbf{\bibinfo{volume}{244}}, \bibinfo{pages}{358} (\bibinfo{year}{1997}).

\bibitem[{\citenamefont{Lang et~al.}(2000)\citenamefont{Lang, Likos, Watzlawek,
  and Lowen}}]{Lang2000Fluid-and-solid}
\bibinfo{author}{\bibfnamefont{A.}~\bibnamefont{Lang}},
  \bibinfo{author}{\bibfnamefont{C.~N.} \bibnamefont{Likos}},
  \bibinfo{author}{\bibfnamefont{M.}~\bibnamefont{Watzlawek}},
  \bibnamefont{and} \bibinfo{author}{\bibfnamefont{H.}~\bibnamefont{Lowen}},
  \bibinfo{journal}{J. Phys.: Condens. Matter} \textbf{\bibinfo{volume}{12}},
  \bibinfo{pages}{5087} (\bibinfo{year}{2000}).

\bibitem[{\citenamefont{Louis et~al.}(2000)\citenamefont{Louis, Bolhuis, and
  Hansen}}]{Louis2000Mean-field-flui}
\bibinfo{author}{\bibfnamefont{A.~A.} \bibnamefont{Louis}},
  \bibinfo{author}{\bibfnamefont{P.~G.} \bibnamefont{Bolhuis}},
  \bibnamefont{and} \bibinfo{author}{\bibfnamefont{J.~P.}
  \bibnamefont{Hansen}}, \bibinfo{journal}{Phys. Rev. E}
  \textbf{\bibinfo{volume}{62}}, \bibinfo{pages}{7961} (\bibinfo{year}{2000}).

\bibitem[{\citenamefont{Prestipino et~al.}(2005)\citenamefont{Prestipino,
  Saija, and Giaquinta}}]{Prestipino2005Phase-diagram-o-JCP}
\bibinfo{author}{\bibfnamefont{S.}~\bibnamefont{Prestipino}},
  \bibinfo{author}{\bibfnamefont{F.}~\bibnamefont{Saija}}, \bibnamefont{and}
  \bibinfo{author}{\bibfnamefont{P.~V.} \bibnamefont{Giaquinta}},
  \bibinfo{journal}{J. Chem. Phys.} \textbf{\bibinfo{volume}{123}},
  \bibinfo{pages}{144110} (\bibinfo{year}{2005}).

\bibitem[{\citenamefont{Giaquinta and
  Saija}(2005)}]{Giaquinta2005Re-entrant-Melt}
\bibinfo{author}{\bibfnamefont{P.~V.} \bibnamefont{Giaquinta}}
  \bibnamefont{and} \bibinfo{author}{\bibfnamefont{F.}~\bibnamefont{Saija}},
  \bibinfo{journal}{Chemphyschem} \textbf{\bibinfo{volume}{6}},
  \bibinfo{pages}{1768} (\bibinfo{year}{2005}).

\bibitem[{\citenamefont{Mausbach and May}(2006)}]{Mausbach2006Static-and-dyna}
\bibinfo{author}{\bibfnamefont{P.}~\bibnamefont{Mausbach}} \bibnamefont{and}
  \bibinfo{author}{\bibfnamefont{H.~O.} \bibnamefont{May}},
  \bibinfo{journal}{Fluid Phase Equilib.} \textbf{\bibinfo{volume}{249}},
  \bibinfo{pages}{17} (\bibinfo{year}{2006}).

\bibitem[{\citenamefont{Zachary et~al.}(2008)\citenamefont{Zachary, Stillinger,
  and Torquato}}]{Zachary2008Gaussian-core-m}
\bibinfo{author}{\bibfnamefont{C.~E.} \bibnamefont{Zachary}},
  \bibinfo{author}{\bibfnamefont{F.~H.} \bibnamefont{Stillinger}},
  \bibnamefont{and} \bibinfo{author}{\bibfnamefont{S.}~\bibnamefont{Torquato}},
  \bibinfo{journal}{J. Chem. Phys.} \textbf{\bibinfo{volume}{128}},
  \bibinfo{pages}{224505} (\bibinfo{year}{2008}).

\bibitem[{\citenamefont{Wensink et~al.}(2008)\citenamefont{Wensink, L\"{o}wen,
  Rex, Likos, and van Teeffelen}}]{Wensink2008Long-time-self-}
\bibinfo{author}{\bibfnamefont{H.}~\bibnamefont{Wensink}},
  \bibinfo{author}{\bibfnamefont{H.}~\bibnamefont{L\"{o}wen}},
  \bibinfo{author}{\bibfnamefont{M.}~\bibnamefont{Rex}},
  \bibinfo{author}{\bibfnamefont{C.}~\bibnamefont{Likos}}, \bibnamefont{and}
  \bibinfo{author}{\bibfnamefont{S.}~\bibnamefont{van Teeffelen}},
  \bibinfo{journal}{Comput. Phys. Commun.} \textbf{\bibinfo{volume}{179}},
  \bibinfo{pages}{77 } (\bibinfo{year}{2008}).

\bibitem[{\citenamefont{Krekelberg et~al.}(2009)\citenamefont{Krekelberg,
  Kumar, Mittal, Errington, and Truskett}}]{Krekelberg2009Gaussian-dynamics}
\bibinfo{author}{\bibfnamefont{W.~P.} \bibnamefont{Krekelberg}},
  \bibinfo{author}{\bibfnamefont{T.}~\bibnamefont{Kumar}},
  \bibinfo{author}{\bibfnamefont{J.}~\bibnamefont{Mittal}},
  \bibinfo{author}{\bibfnamefont{J.~R.} \bibnamefont{Errington}},
  \bibnamefont{and} \bibinfo{author}{\bibfnamefont{T.~M.}
  \bibnamefont{Truskett}}, \bibinfo{journal}{Physical Review E}
  \textbf{\bibinfo{volume}{79}}, \bibinfo{pages}{031203}
  (\bibinfo{year}{2009}).

\bibitem[{\citenamefont{Stillinger}(1976)}]{Stillinger1976Phase-transitio}
\bibinfo{author}{\bibfnamefont{F.~H.} \bibnamefont{Stillinger}},
  \bibinfo{journal}{J. Chem. Phys.} \textbf{\bibinfo{volume}{65}},
  \bibinfo{pages}{3968} (\bibinfo{year}{1976}).

\bibitem[{\citenamefont{Likos}(2001)}]{Likos2001effective-interactions}
\bibinfo{author}{\bibfnamefont{C.~N.} \bibnamefont{Likos}},
  \bibinfo{journal}{Physics Reports} \textbf{\bibinfo{volume}{348}},
  \bibinfo{pages}{267 } (\bibinfo{year}{2001}).

\bibitem[{\citenamefont{Stillinger and
  Weber}(1978)}]{Stillinger1978Study-of-meltin}
\bibinfo{author}{\bibfnamefont{F.~H.} \bibnamefont{Stillinger}}
  \bibnamefont{and} \bibinfo{author}{\bibfnamefont{T.~A.} \bibnamefont{Weber}},
  \bibinfo{journal}{J. Chem. Phys.} \textbf{\bibinfo{volume}{68}},
  \bibinfo{pages}{3837} (\bibinfo{year}{1978}).

\bibitem[{\citenamefont{Rosenfeld}(1977)}]{Rosenfeld1977Relation-betwee}
\bibinfo{author}{\bibfnamefont{Y.}~\bibnamefont{Rosenfeld}},
  \bibinfo{journal}{Phys. Rev. A} \textbf{\bibinfo{volume}{15}},
  \bibinfo{pages}{2545} (\bibinfo{year}{1977}).

\bibitem[{\citenamefont{Rosenfeld}(1999)}]{Rosenfeld1999A-quasi-univers}
\bibinfo{author}{\bibfnamefont{Y.}~\bibnamefont{Rosenfeld}},
  \bibinfo{journal}{J. Phys.: Condens. Matter} \textbf{\bibinfo{volume}{11}},
  \bibinfo{pages}{5415} (\bibinfo{year}{1999}).

\bibitem[{\citenamefont{Dzugutov}(1996)}]{Dzugutov1996A-univeral-scal}
\bibinfo{author}{\bibfnamefont{M.}~\bibnamefont{Dzugutov}},
  \bibinfo{journal}{Nature} \textbf{\bibinfo{volume}{381}},
  \bibinfo{pages}{137} (\bibinfo{year}{1996}).

\bibitem[{\citenamefont{Mittal et~al.}(2007)\citenamefont{Mittal, Errington,
  and Truskett}}]{Mittal2007Relationships-b}
\bibinfo{author}{\bibfnamefont{J.}~\bibnamefont{Mittal}},
  \bibinfo{author}{\bibfnamefont{J.}~\bibnamefont{Errington}},
  \bibnamefont{and} \bibinfo{author}{\bibfnamefont{T.}~\bibnamefont{Truskett}},
  \bibinfo{journal}{J. Phys. Chem. B} \textbf{\bibinfo{volume}{111}},
  \bibinfo{pages}{10054} (\bibinfo{year}{2007}).

\bibitem[{\citenamefont{Errington and
  Debenedetti}(2001)}]{Errington2001Relationship-be}
\bibinfo{author}{\bibfnamefont{J.~R.} \bibnamefont{Errington}}
  \bibnamefont{and} \bibinfo{author}{\bibfnamefont{P.~G.}
  \bibnamefont{Debenedetti}}, \bibinfo{journal}{Nature}
  \textbf{\bibinfo{volume}{409}}, \bibinfo{pages}{318} (\bibinfo{year}{2001}).

\bibitem[{\citenamefont{Mittal et~al.}(2006{\natexlab{a}})\citenamefont{Mittal,
  Errington, and Truskett}}]{Mittal2006Relationship-be}
\bibinfo{author}{\bibfnamefont{J.}~\bibnamefont{Mittal}},
  \bibinfo{author}{\bibfnamefont{J.~R.} \bibnamefont{Errington}},
  \bibnamefont{and} \bibinfo{author}{\bibfnamefont{T.~M.}
  \bibnamefont{Truskett}}, \bibinfo{journal}{J. Chem. Phys.}
  \textbf{\bibinfo{volume}{125}}, \bibinfo{pages}{076102}
  (\bibinfo{year}{2006}{\natexlab{a}}).

\bibitem[{\citenamefont{Errington et~al.}(2006)\citenamefont{Errington,
  Truskett, and Mittal}}]{Errington2006Excess-entropy-}
\bibinfo{author}{\bibfnamefont{J.~R.} \bibnamefont{Errington}},
  \bibinfo{author}{\bibfnamefont{T.~M.} \bibnamefont{Truskett}},
  \bibnamefont{and} \bibinfo{author}{\bibfnamefont{J.}~\bibnamefont{Mittal}},
  \bibinfo{journal}{J. Chem. Phys.} \textbf{\bibinfo{volume}{125}},
  \bibinfo{pages}{244502} (\bibinfo{year}{2006}).

\bibitem[{\citenamefont{Mittal et~al.}(2006{\natexlab{b}})\citenamefont{Mittal,
  Errington, and Truskett}}]{Mittal2006Quantitative-Li}
\bibinfo{author}{\bibfnamefont{J.}~\bibnamefont{Mittal}},
  \bibinfo{author}{\bibfnamefont{J.~R.} \bibnamefont{Errington}},
  \bibnamefont{and} \bibinfo{author}{\bibfnamefont{T.~M.}
  \bibnamefont{Truskett}}, \bibinfo{journal}{J. Phys. Chem. B}
  \textbf{\bibinfo{volume}{110}}, \bibinfo{pages}{18147}
  (\bibinfo{year}{2006}{\natexlab{b}}).

\bibitem[{\citenamefont{Krekelberg et~al.}(2007)\citenamefont{Krekelberg,
  Mittal, Ganesan, and Truskett}}]{Krekelberg2007How-short-range}
\bibinfo{author}{\bibfnamefont{W.~P.} \bibnamefont{Krekelberg}},
  \bibinfo{author}{\bibfnamefont{J.}~\bibnamefont{Mittal}},
  \bibinfo{author}{\bibfnamefont{V.}~\bibnamefont{Ganesan}}, \bibnamefont{and}
  \bibinfo{author}{\bibfnamefont{T.~M.} \bibnamefont{Truskett}},
  \bibinfo{journal}{J. Chem. Phys.} \textbf{\bibinfo{volume}{127}},
  \bibinfo{pages}{044502} (\bibinfo{year}{2007}).

\bibitem[{\citenamefont{Archer and Evans}(2001)}]{Archer2001Binary-Gaussian}
\bibinfo{author}{\bibfnamefont{A.~J.} \bibnamefont{Archer}} \bibnamefont{and}
  \bibinfo{author}{\bibfnamefont{R.}~\bibnamefont{Evans}},
  \bibinfo{journal}{Phys. Rev. E} \textbf{\bibinfo{volume}{64}},
  \bibinfo{pages}{041501} (\bibinfo{year}{2001}).

\bibitem[{\citenamefont{Allen and Tildesley}(1987)}]{Allen1987Computer-Simula}
\bibinfo{author}{\bibfnamefont{M.~P.} \bibnamefont{Allen}} \bibnamefont{and}
  \bibinfo{author}{\bibfnamefont{D.~J.} \bibnamefont{Tildesley}},
  \emph{\bibinfo{title}{Computer Simulations of Liquids}}
  (\bibinfo{publisher}{Oxford University Press, New York},
  \bibinfo{year}{1987}).

\bibitem[{not({\natexlab{a}})}]{note-error-est}
\bibinfo{note}{We use Student's t distribution together with the tracer
  diffusivities from five independent runs to estimate 95\% confidence
  intervals for the $D_i$.}

\bibitem[{\citenamefont{Hernando}(1990)}]{Hernando1990Thermodynamic-pot}
\bibinfo{author}{\bibfnamefont{J.}~\bibnamefont{Hernando}},
  \bibinfo{journal}{Molecular Physics} \textbf{\bibinfo{volume}{69}},
  \bibinfo{pages}{319} (\bibinfo{year}{1990}).

\bibitem[{\citenamefont{Samanta et~al.}(2001)\citenamefont{Samanta, Ali, and
  Ghosh}}]{Samanta2001Universal-Scali}
\bibinfo{author}{\bibfnamefont{A.}~\bibnamefont{Samanta}},
  \bibinfo{author}{\bibfnamefont{S.~M.} \bibnamefont{Ali}}, \bibnamefont{and}
  \bibinfo{author}{\bibfnamefont{S.~K.} \bibnamefont{Ghosh}},
  \bibinfo{journal}{Phys. Rev. Lett.} \textbf{\bibinfo{volume}{87}},
  \bibinfo{pages}{245901} (\bibinfo{year}{2001}).

\bibitem[{\citenamefont{Truskett et~al.}(2000)\citenamefont{Truskett, Torquato,
  and Debenedetti}}]{Truskett2000Towards-a-quant}
\bibinfo{author}{\bibfnamefont{T.~M.} \bibnamefont{Truskett}},
  \bibinfo{author}{\bibfnamefont{S.}~\bibnamefont{Torquato}}, \bibnamefont{and}
  \bibinfo{author}{\bibfnamefont{P.~G.} \bibnamefont{Debenedetti}},
  \bibinfo{journal}{Phys. Rev. E} \textbf{\bibinfo{volume}{62}},
  \bibinfo{pages}{993} (\bibinfo{year}{2000}).

\bibitem[{not({\natexlab{b}})}]{note-anomalies}
\bibinfo{note}{In the limit of pure B ($x_A=0$), one
  observes\cite{Krekelberg2009Gaussian-dynamics} both $(\partial D_B / \partial
  \rho)_{T,x_A}>0$ and $(\partial \stwoi[B] / \partial \rho)_{T,x_A}>0$ for
  reduced densities greater than $\rho \sigma_{AA}^3 \approx 1.4$. However, we
  have found in our studies of this system that, for non-zero values of $x_A$,
  the binary GC fluid will often phase separate at densities below where the
  onset of anomalous dynamic and structural behavior occurs.}

\end{thebibliography}

\end{document}